\documentstyle[epsfig, twocolumn]{esapub}

\begin{document}

%% Do not remove the following six lines:
\setlength{\parindent}{0pt}
\setlength{\parskip}{ 10pt plus 1pt minus 1pt}
\setlength{\hoffset}{-1.5truecm}
\setlength{\textwidth}{ 17.1truecm }
\setlength{\columnsep}{1truecm }
\setlength{\columnseprule}{0pt}
\setlength{\headheight}{12pt}
\setlength{\headsep}{20pt}
\pagestyle{esapubheadings}

%% Title - should be in capitals:
\title{\bf ULTRAVIOLET ISOTROPIC EMISSION FROM THE BLAZAR 3C 279}
%\title{\bf A \LaTeX{}-BASED FORMAT FOR PROCEEDINGS\thanks{these instructions
%        should be used to prepare manuscripts for the published proceedings}}

%% If the author list spans more than one line then the {\bf (bold
%% font)} command must be inserted for each line
\author{{\bf E. Pian$^{1,2}$, A. Koratkar$^2$, L. Maraschi$^3$, C. M. Urry$^2$, 
G. Madejski$^4$,} \vspace{2mm} \\
{\bf I. M. McHardy$^5$, J. E. Pesce$^2$, A. Treves$^6$, P. Grandi$^7$,
C. M. Leach$^5$} \vspace{2mm}\\
$^1$ ITESRE-CNR, Via Gobetti 101, I-40129 Bologna, Italy \\
$^2$ STScI, 3700 San Martin Drive, Baltimore, MD 21218, USA \\
$^3$ Brera Astr. Obs., Via Brera 28, I-20121 Milan, Italy \\
$^4$ LHEA/GSFC, Greenbelt, MD 20771, USA \\
$^5$ Dpt. of Phys., Univ. of Southampton, Southampton SO9 5NH, UK \\
$^6$ Dpt. of Phys., Univ. of Milan, Via Lucini 3, I-22100 Como, Italy \\
$^7$ IAS-CNR, Via Fosso del Cavaliere, I-00133 Roma, Italy}

\maketitle

\begin{abstract}

Archival IUE SWP and HST FOS spectra show the presence of a 
relatively strong,
broad Lyman $\alpha$ emission line superposed onto the UV spectral 
continuum of the blazar 3C~279. As opposed to a factor $\sim$50 
variation of the continuum flux during eight years, the emission 
line did not exhibit significant intensity changes.
Simultaneous IUE SWP and LWP spectra of 3C~279 in low emission
state are fitted by power-laws of index $\alpha_\nu \sim 1$ 
($f_\nu \propto \nu^{-\alpha_\nu}$), significantly flatter than 
measured during higher states.  Our observations suggest that the
Lyman $\alpha$ line is not powered by the beamed, anisotropic 
synchrotron radiation which produces the observed continuum in 
3C~279, but rather by an unbeamed component characterized by slower 
and lower amplitude variability. The latter may account for
the UV continuum observed during the
very low state of January 1993.
\vspace {5pt} \\

%% Do not remove the previous commands. Your abstract should 
%% end with \vspace {5pt} \\  

%% Please insert your keywords here.
  Key~words: ultraviolet spectra; blazar emission lines; blazar emission
mechanisms; accretion disks.

\end{abstract}

\section{INTRODUCTION}

The blazar 3C~279 ($z = 0.54$) is well studied and shows frequent large 
continuum
flares from radio to gamma-ray wavelengths. Inverse Compton scattering
of relativistic electrons off synchrotron or ambient photons is likely
responsible for the emission at hard X- and gamma-ray energies. 
Clarifying the exact nature of the seed photons for this mechanism 
would explain the origin of the huge amplitude variations exhibited
by 3C~279 at the highest energies (Maraschi et al. 1994; Hartman et al. 
1996; Wehrle et al. 1998).
There have been several multi-wavelength observations of this blazar, 
and hence there are many UV data available in the archives. 
In particular, 3C~279 was monitored on a nearly daily basis with IUE 
and ROSAT for three weeks between December 1992 and January 1993, 
simultaneously with gamma-ray observations by EGRET, and with coordinated
optical observations.  
We present here a study of the correlated variability of the UV
continuum and of the broad Lyman $\alpha$ emission line intensity over
eight years, and
compare our results with the findings of simultaneous optical, UV and X-ray
monitoring in the period 2-5 January 1993.

\section{DATA ANALYSIS AND RESULTS}

IUE SWP and LWP spectra have been extracted with the NEWSIPS routine
used for the implementation of the IUE Final Archive (Nichols \&
Linsky 1996) and
dereddened using a neutral hydrogen column
density $N_{HI} = 2.22 \times 10^{20}$ cm$^{-2}$ (Elvis et al. 1989).
The continuum flux at 1750 \AA\ has been derived through
a power-law fit to the SWP spectra;
line intensities have been
calculated with a Gaussian fit (see Koratkar et al. 1998 for
details).
To increase the signal-to-noise ratio of the Lyman $\alpha$ 
emission line, we binned the
spectra in four groups depending on the continuum flux at 1750 \AA. The
SWP spectra in each group were then co-added, as follows:  
`high spectrum': $f_{cont} >$ 
$2 \times 10^{-14}$ erg cm$^{-2}$ s$^{-1}$ \AA$^{-1}$,
`medium spectrum': $8\times10^{-15} < f_{cont} \leq 2\times10^{-14}$ 
erg cm$^{-2}$ s$^{-1}$ \AA$^{-1}$, 
`low spectrum': $3\times10^{-15} < f_{cont} \leq 8\times10^{-15}$ 
erg cm$^{-2}$ s$^{-1}$ \AA$^{-1}$, 
 `very low spectrum': $1\times10^{-15} \leq f_{cont} \leq 3\times10^{-15}$ 
erg cm$^{-2}$ s$^{-1}$ \AA$^{-1}$.  

HST FOS spectra in 1992, 1994, and 1996 have been retrieved from the 
archive, analyzed with the STSDAS/IRAF reduction package
and dereddened (see Koratkar et al. 1998).

ROSAT PSPC spectra were taken from 1992 December 27 to 1993 January 
13 on a daily basis.  The background,
carefully chosen in a region free from faint field sources, 
was subtracted from the data. 
The neutral hydrogen column density which
yields the best power-law fit for all the spectra is $2.5 \times 10^{20}$
cm$^{-2}$, which is consistent with Galactic within  the 10\% 1-$\sigma$ 
error. Fixing $N_{HI}$ at this value, good fits to the spectra are obtained 
with single power-laws of index $0.7 \div 0.8$.  
More details will be reported in a forthcoming paper (Pian et al. 1998).

In all UV continuum flux ranges the Lyman $\alpha$ 
emission line is visible redshifted to $\sim$1868 \AA\ (Koratkar et al. 
1998).
Its strength  is nearly constant
($\sim$5 $\times 10^{-14}$ erg~cm$^{-2}$ s$^{-1}$), while the
continuum at 1750~\AA\ varies by a factor of $\sim$50, from $\sim 0.6$ to
31.6 $\times 10^{-15}$ erg~cm$^{-2}$ s$^{-1}$ \AA$^{-1}$ (Fig. 1).
The emission line equivalent width ranges between 1 and 45 \AA\ (Fig. 2).

\begin{figure}[h]
  \begin{center}
    \leavevmode
  \centerline{\epsfig{file=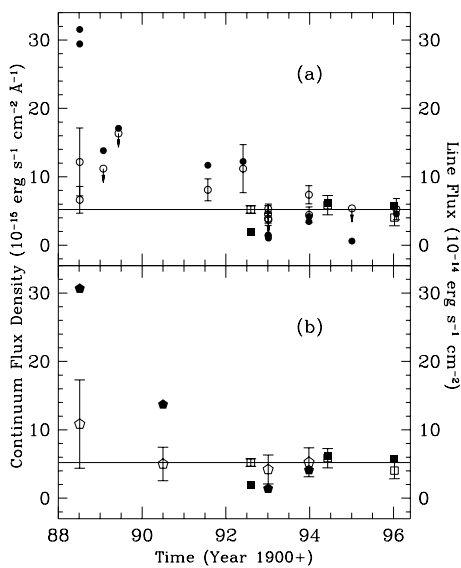,width=6cm}}
%  \vspace{1.cm}
  \end{center}
  \caption{\em (a) The continuum flux density (filled points) and
Lyman $\alpha$ line flux (open points) of 3C~279 as a function of time.
The IUE data are represented by circles and the HST/FOS data by
squares. Error bars are smaller than the symbol if not indicated in the
figure. (b) Same as (a) but for co-added IUE spectra (pentagons) and 
HST/FOS data (squares).}
%\label{fig:sample1}
\end{figure}

\begin{figure}[h]
  \begin{center}
    \leavevmode
  \centerline{\epsfig{file=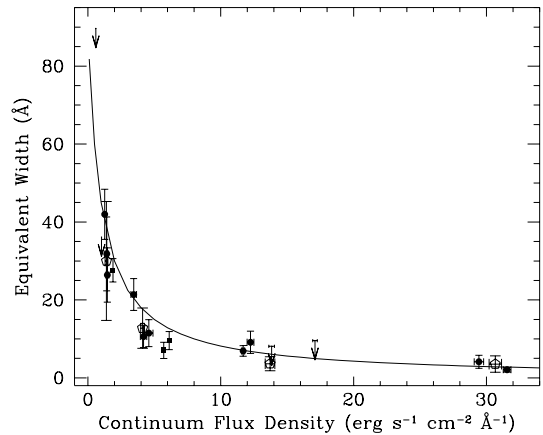,width=6cm}}
%  \vspace{1.cm}
  \end{center}
  \caption{\em The Lyman $\alpha$ line equivalent width as a function
of the continuum flux density (circles, IUE individual;
pentagons, IUE co-added; squares, HST/FOS). The solid line represents the
variation in equivalent width for a constant line intensity and a varying
beamed continuum.}

%\label{fig:sample1}
\end{figure}

In 2-5 January 1993, the UV emission of 3C~279 was at a historical 
minimum in the whole IUE range (SWP and LWP).
Power-laws have been fitted to simultaneous pairs of SWP and
LWP dereddened spectra.  
The average spectrum in the 1200-2700 \AA\ interval is 
described by a spectral index $\alpha_\nu \simeq 1$, which is
unusually flat for this object, as apparent from a comparison with
spectra taken at epochs of higher UV flux level (Fig. 3),
and harder than the simultaneous optical spectrum (Fig. 4).

\begin{figure*}[h]
  \begin{center}
    \leavevmode
  \centerline{\epsfig{file=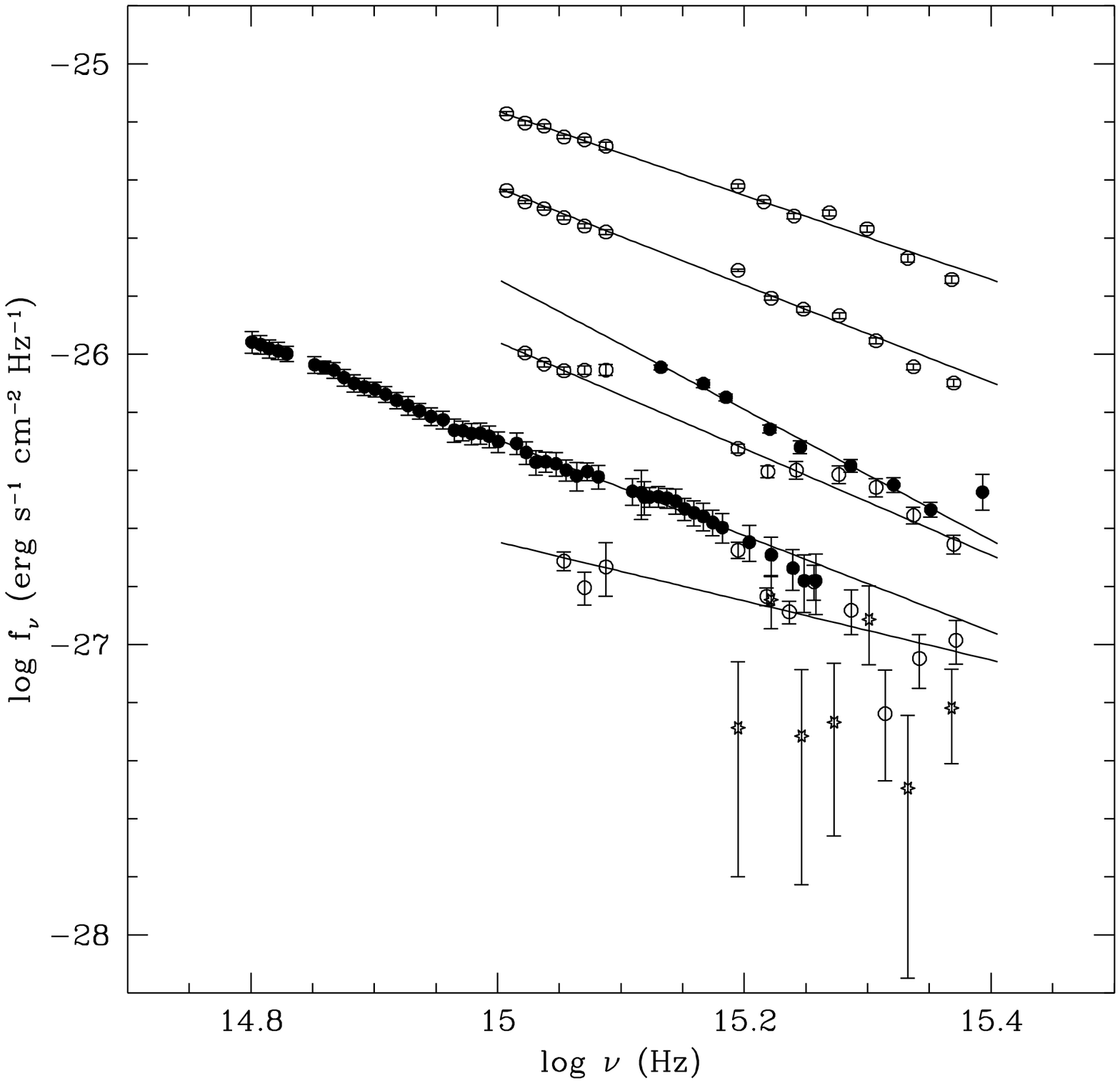,width=8cm}}
%  \vspace{1.cm}
  \end{center}
  \caption{\em Dereddened and binned UV spectra with their power-law
best fits.  The IUE continua
(open symbols) have been co-added according to the flux level.
Spectral indices are $\alpha = 1.45 \pm 0.04$ (high state), 
$\alpha = 1.68 \pm 0.03$ (medium), 
$\alpha = 1.84 \pm 0.08$ (low state), 
$\alpha = 1.0 \pm 0.2$ (very low state). 
An extremely low state IUE-SWP spectrum is also shown (Jan 1995, stars), 
but no fit has been attempted, due to the poor S/N. HST spectra
(filled dots) refer to April 1992 (lower state, $\alpha = 1.7 \pm 0.2$)
and January 1996 (higher state, $\alpha = 2.25 \pm 0.06$).}

\end{figure*}

\section{DISCUSSION}

In eight years, the UV continuum of 3C~279 has varied by a 
factor $\sim$50, while the Lyman $\alpha$ line flux has remained 
nearly
constant.  This suggests that the observed highly variable continuum,
 most likely due
to beamed  synchrotron radiation from a relativistic jet
 does not contribute significantly in powering  the emission line. 

In fact the line equivalent width in 3C 279 is smaller than
observed in 'normal' quasars, where the observed continuum, 
probably thermal radiation from an accretion disk, is also   
the source of ionizing radiation. 
If the line emitting gas in the broad line region of 3C 279 is 
also ionized by an inner disk, its radiation, usually swamped 
by the beamed blazar continuum, may become observable in 
very low states.

\begin{figure*}[h]
  \begin{center}
    \leavevmode
  \centerline{\epsfig{file=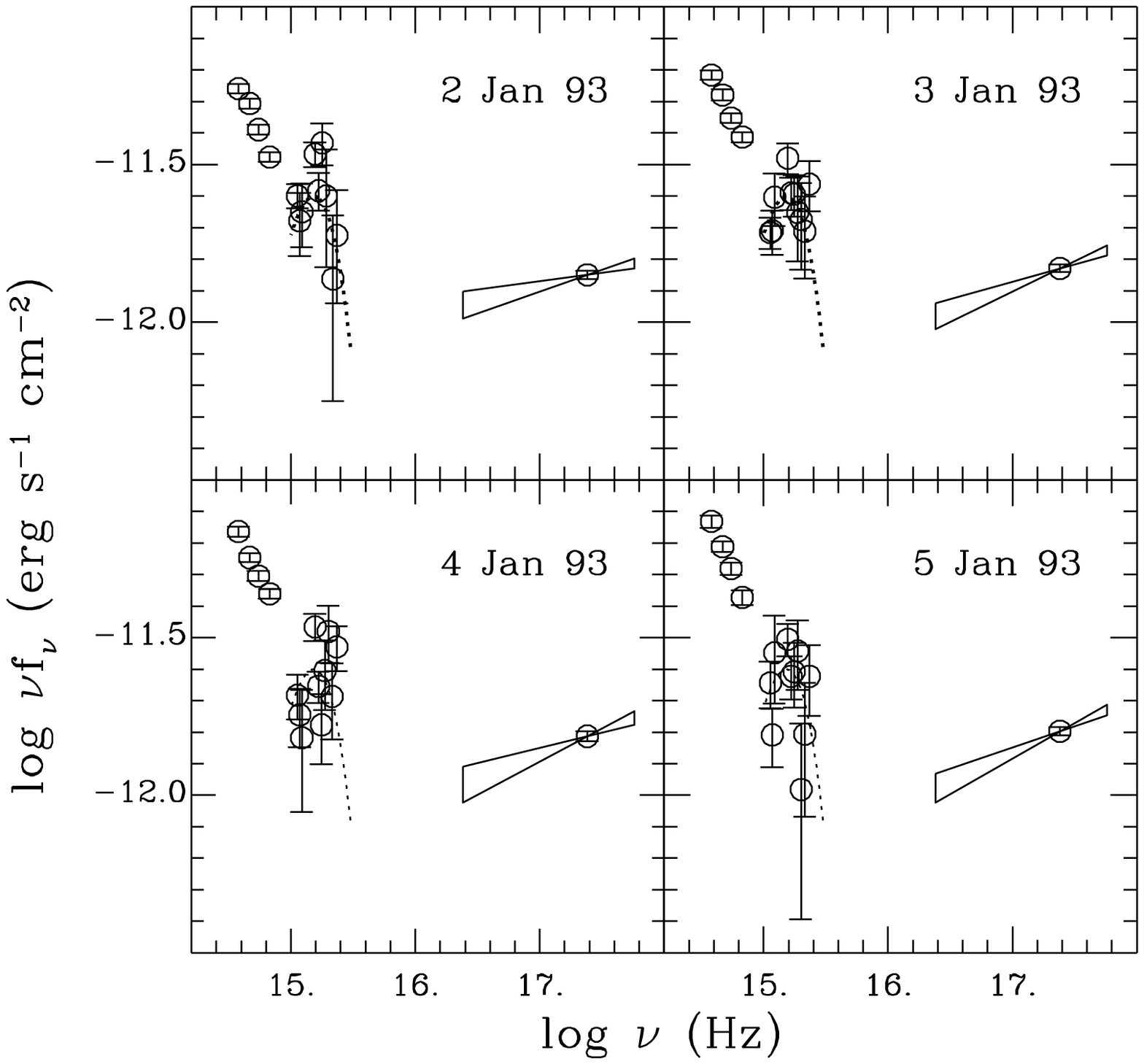,width=10cm}}
%  \vspace{1.cm}
  \end{center}
  \caption{\em De-extincted simultaneous optical (from Grandi et
al. 1996), UV (IUE) and soft X-ray (ROSAT)
energy distributions in 2-5 January 1993.  The power-law fits to
the ROSAT data and the black body fit to the IUE LWP and SWP data
are shown as solid and dashed curves, respectively
(1-$\sigma$ ranges are also reported for the power-laws).}

%\label{fig:sample1}
\end{figure*}

The spectral flattening observed in the 1200-2700 \AA\ range 
in correspondence with a very low UV continuum level is an 
uncommon feature
in blazars, which generally exhibit spectral hardening during
brighter states, and might represent the 
signature of the putative thermal, isotropic component underlying 
the highly variable, beamed continuum of 3C~279, and photoionizing 
the line emitting gas (Fig. 4).   The low state soft X-ray 
spectrum, which is well
described ($\chi^2 \sim 1$) by a power-law
of index $\alpha_\nu \sim 0.7-0.8$ may also contain a 
Seyfert-like component.
Assuming a simple accretion disk model described by a 
single black body, this would have a temperature
of $\sim$20000 K and a size of $\sim$1 light day.
The observation of a still weaker flux in January 
1995 at the shorter UV
wavelengths (Fig. 3) suggests some variability
of this thermal component, albeit modest.
More sensitive observations in the 
hard X-ray / gamma-ray band during a low state
would be needed to constrain this hypothesis.

The presence of intense line emission and the suggestion of
a thermal component in 3C~279 are consistent with the
scenario in which the seed photons for the inverse Compton 
mechanism producing 
the gamma-rays are external to the relativistic jet and provided
either by the broad line region or by the inner accretion disk.
 However, the observed  large amplitude variability in gamma-rays
accompanied by lower amplitude variability at lower energies
requires not only changes in the energetic electrons in the jet,
{\it but also variations in the soft photon field}, at least in 
a simple one zone
emission model. A possible scenario is proposed  by Ghisellini 
and Madau (1996),
whereby some line emission is induced by radiation from the jet. 
It is interesting to ask whether this mechanism would have 
some observable consequences on the observed line emission.

Although intriguing and promising, our results on the UV continuum 
and Lyman $\alpha$ line characteristics in 3C~279 must be taken with
caution: 
the available IUE and HST spectra are too few and too sparse in time
to yield a definitive proof of lack of correlated variability between 
continuum and emission line at the shorter (one day or less) 
time scales.  Moreover, 
the low signal-to-noise ratio of the data prevent a very accurate
measurement of the spectral index in low UV emission state.

Further data are necessary to confirm our findings. An intensive
and regular monitoring of the UV spectrum of 3C~279 and other 
blazars, along with a detailed
sampling during low emission states would clarify the existence and role 
of an isotropic emission component in this class of active galactic
nuclei, and possibly lead to a link between blazar and normal
quasar properties.  This task can be pursued more favourably in
the UV spectral range, where the high ionization emission lines
are located (at low or intermediate redshift) and which is less
diluted by the stellar contribution of the host galaxy. 
This open problem represents an important heritage of IUE and its
solution could be attempted only by a UV observing facility with
its same easy and flexible scheduling.

%\section{REFERENCES}

\end{document}